# New infrared spectra of $CO_2$ – Ne: fundamental for $CO_2$ –$^{22}$Ne isotopologue and symmetry breaking of the intramolecular $CO_2$ bend

A.J. Barclay,[1] A.R.W. McKellar,[2] and N. Moazzen-Ahmadi[1]

[1] *Department of Physics and Astronomy, University of Calgary, 2500 University Drive North West, Calgary, Alberta T2N 1N4, Canada*

[2] *National Research Council of Canada, Ottawa, Ontario K1A 0R6, Canada*

**Abstract**

The infrared spectrum of the weakly-bound $CO_2$-Ne complex is studied in the region of the carbon dioxide $\nu_3$ fundamental vibration ($\approx$2350 cm$^{-1}$), using a tunable OPO laser source to probe a pulsed supersonic slit jet expansion. For the fundamental $CO_2$ transition $(\nu_1, \nu_2^{l2}, \nu_3) = (00^01) \leftarrow (00^00)$, both $CO_2$-$^{20}$Ne and $CO_2$-$^{22}$Ne are assigned and analyzed in combination with available microwave data to obtain the best currently available molecular parameters. For the hot band $CO_2$ transition, $(01^11) \leftarrow (01^10)$, detection of the weak $CO_2$-Ne spectrum reveals the symmetry breaking of the $CO_2$ $\nu_2$ bending mode induced by the Ne atom, with the out-of-plane component determined to lie 0.057 cm$^{-1}$ higher in energy than the in-plane component.

1. **Introduction**

The simplicity of the $CO_2$-rare gas clusters makes them ideal as the probe of the angular anisotropy of the intermolecular potential and test cases for improved theoretical models with the goal of rigorous fully dimensional and fully coupled intra- and inter-molecular rovibrational calculation, which we hope is realized soon. This task becomes amenable after the development of full-dimensional potential energy surface, 5D in the case of $CO_2$-rare gases. Towards this end, we have made new infrared observations for $CO_2$-Ar,[1] $CO_2$-Xe [2] and $CO_2$-Kr.[3] Here we report new IR observations for $CO_2$-Ne.

The first high-resolution spectroscopic results on the weakly-bound $CO_2$-Ne complex were reported in 1988 by Randall et al.[4] and by Fraser et al.[5] In the former paper, the infrared spectrum of the dimer was observed in the region of the $\nu_3$ fundamental band of $CO_2$ near 2350 cm$^{-1}$. In the latter paper, pure rotational microwave spectra were observed as well as infrared spectra in the $CO_2$ $\nu_1 + \nu_3$ (3710 cm$^{-1}$) and $2\nu_2 + \nu_3$ (3610 cm$^{-1}$) regions. Subsequently, there has been a more extensive microwave study[6] of $CO_2$-Ne involving a number of isotopologues, and a further infrared study[7] involving the $\nu_3$ band of $^{12}C^{18}O_2$ (2314 cm$^{-1}$). On the theoretical side, a couple of detailed potential energy surfaces have been reported for the $CO_2$-Ne interaction.[8,9] The latter paper includes the dependence of the potential on the $CO_2$ $\nu_3$ (asymmetric stretch) vibration in order to better represent infrared spectra, and further results using this potential were reported in two follow-up papers.[10,11]

The minimum energy structure of $CO_2$-Ne (and the other $CO_2$-Rg dimers) is T-shaped, having the Ne atom located "beside" the linear $CO_2$ molecule with an effective C to Ne distance of about 3.29 Å. As a result, the *a*-inertial axis connects C and Ne, the *b*-axis is parallel to the O-

C-O axis, and the *c*-axis is perpendicular to the $CO_2$-Ne plane. Nuclear spin statistics allow only even values of $K_a$ for the ground state of dimers containing $C^{16}O_2$ (or $C^{18}O_2$).

In the present paper, we reexamine the spectrum of $CO_2$-Ne in the $CO_2$ $\nu_3$ region. Coverage of the fundamental band of $CO_2$-Ne is extended, and $CO_2$-$^{22}$Ne is observed in natural abundance and analyzed. A weak spectrum of $CO_2$-Ne is also detected in the region of the $CO_2$ $(01^11) - (01^10)$ hot band near 2337 cm$^{-1}$. This provides a determination of the splitting of the degenerate $CO_2$ $\nu_2$ bending vibration into two modes (in-plane and out-of-plane) induced by the presence of the nearby Ne atom.

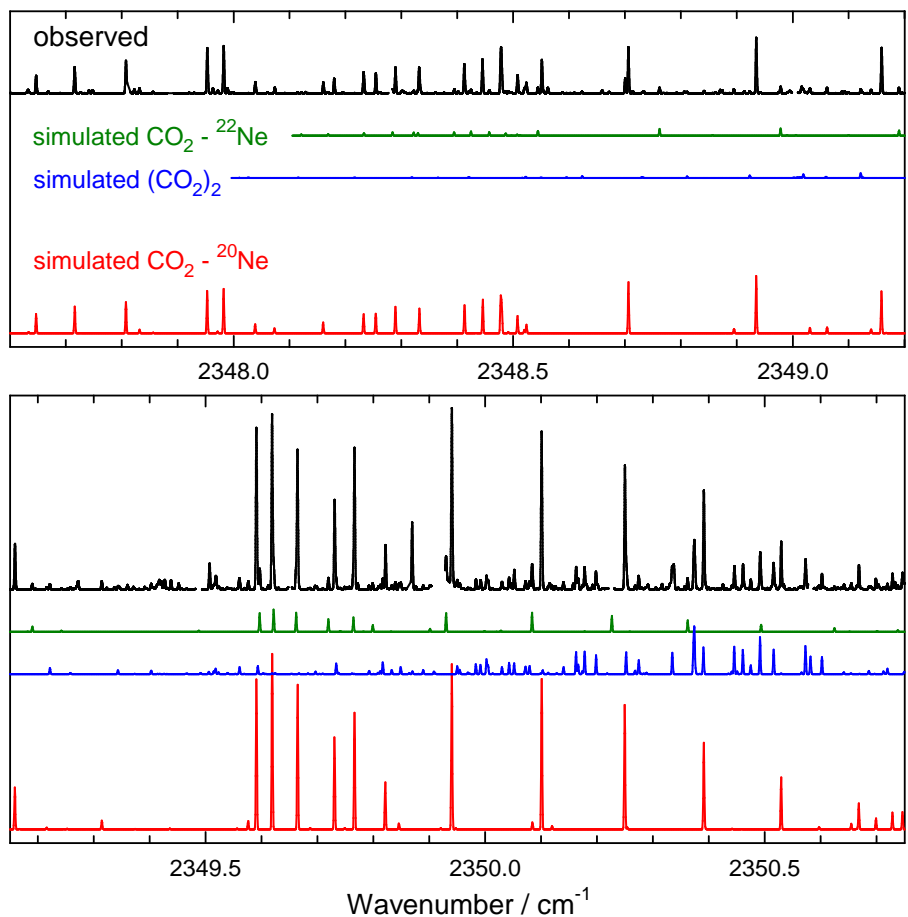

Fig. 1. Observed and simulated (T = 2 K) spectra of $CO_2$-Ne in the region of the $CO_2$ $\nu_3$ band. Gaps in the observed spectrum correspond to regions of $CO_2$ monomer or $CO_2$-He absorption, and the simulation also shows $CO_2$ dimer, which becomes significant above about 2350 cm$^{-1}$.

Table 1. Molecular parameters for the fundamental band of $CO_2$-Ne (in cm$^{-1}$) [a]

|  | $CO_2$-$^{20}$Ne Ground State | $CO_2$-$^{20}$Ne Excited State | $CO_2$-$^{22}$Ne Ground State | $CO_2$-$^{22}$Ne Excited State |
|---|---|---|---|---|
| $\nu_0$ |  | 2349.2796(1) |  | 2349.2819(1) |
| $A$ | 0.402339 (19) | 0.399171(17) | 0.402061(12) | 0.398786(10) |
| $B$ | 0.11537499(75) | 0.1151810(28) | 0.10831698(41) | 0.1081390(24) |
| $C$ | 0.08772499(76) | 0.0874331(18) | 0.08358319(42) | 0.0833061(21) |
| $10^5 \times \Delta_K$ | -6.39(14) | [-6.39] | -7.592 (81) | [-7.592] |
| $10^5 \times \Delta_{JK}$ | 7.36430(78) | [7.36430] | 6.56697(38) | [6.56697] |
| $10^6 \times \Delta_J$ | 4.5224(48) | [4.5224] | 4.0004(24) | [4.0004] |
| $10^5 \times \delta_K$ | 5.2279(104) | [5.2279] | 4.6096(58) | [4.6096] |
| $10^6 \times \delta_J$ | 1.0605(53) | [1.0605] | 0.8893(29) | [0.8893] |
| $10^{10} \times H_J$ | -10.0(17) | [-10.0] | -7.28(84) | [-7.28] |

[a] Quantities in parentheses correspond to 1σ from the least-squares fit, in units of the last quoted digit. The ground and excited state centrifugal distortion parameters were constrained to be equal.

## 2. Results

Spectra were recorded as described previously.[12-14] A pulsed supersonic slit jet expansion was probed by a rapid-scan optical parametric oscillator source. The gas expansion mixture contained about 0.04% carbon dioxide plus 0.9% neon in helium carrier gas with a jet backing pressure of about 13 atmospheres. Wavenumber calibration was carried out by simultaneously recording signals from a fixed etalon and a $CO_2$ reference gas cell. Simulation and fitting were carried out using PGOPHER software.[15]

### 2.1. Fundamental band, $CO_2$ $(00^01) \leftarrow (00^00)$

The observed spectrum in the central part of the fundamental band is shown in Fig. 1. This is a perpendicular $b$-type band ($\Delta K_a = \pm 1$) with only $K_a$ = even levels in the ground state and $K_a$ = odd in the excited state. In Fig. 1 we see $K_a = 1 \leftarrow 0$ and $1 \leftarrow 2$ subbands with $Q$-branches at about 2349.6 and 2348.4 cm$^{-1}$, respectively. Further below and above the region shown in the figure, we also observed $K_a = 3 \leftarrow 4$ and $3 \leftarrow 2$ transitions. The natural abundance of $^{22}$Ne is about 9%, and it was reasonably straightforward to assign some transitions of $CO_2$-$^{22}$Ne in addition to those of the dominant $CO_2$-$^{20}$Ne isotopologue. Ultimately we assigned 68 transitions of $CO_2$-$^{20}$Ne, and 32 of $CO_2$-$^{22}$Ne, which were analyzed to obtain the parameters listed in Table 1. The analyses also included 8 pure rotational microwave transitions for each isotopologue, taken from the paper by Xu and Jäger[6] and weighted to reflect their higher precision. These microwave data essentially determined the ground state parameters except for $A$, to which they are not very sensitive. Excellent fits were obtained by constraining the ground and excited state centrifugal distortion parameters to be equal. The infrared root mean square errors were 0.00019 and 0.00010 cm$^{-1}$, for $CO_2$-$^{20}$Ne and $CO_2$-$^{22}$Ne, respectively, and the microwave rms errors were 1.2 and 0.1 kHz. Observed and calculated line positions are given as Supplementary Information.

The parameters for $CO_2$-$^{20}$Ne in Table 1 agree quite well with those of Randall et al.,[4] but they are more accurate thanks to the wider range of infrared data and the inclusion of microwave data. The parameters for $CO_2$-$^{22}$Ne are new, particularly those for the excited state. We see that the band origin of $CO_2$-$^{22}$Ne is slightly (0.002 cm$^{-1}$) higher than that of $CO_2$-$^{20}$Ne, so that the vibrational blue shift relative to the free $CO_2$ molecule increases from 0.1363 to 0.1386 cm$^{-1}$. This change can be explained by noting that the average intermolecular distance is slightly smaller for $CO_2$-$^{22}$Ne due to the anharmonicity of the van der Waals bond. This allows $^{22}$Ne to get slightly closer to $CO_2$ which induces a larger shift in the $CO_2$ vibration.

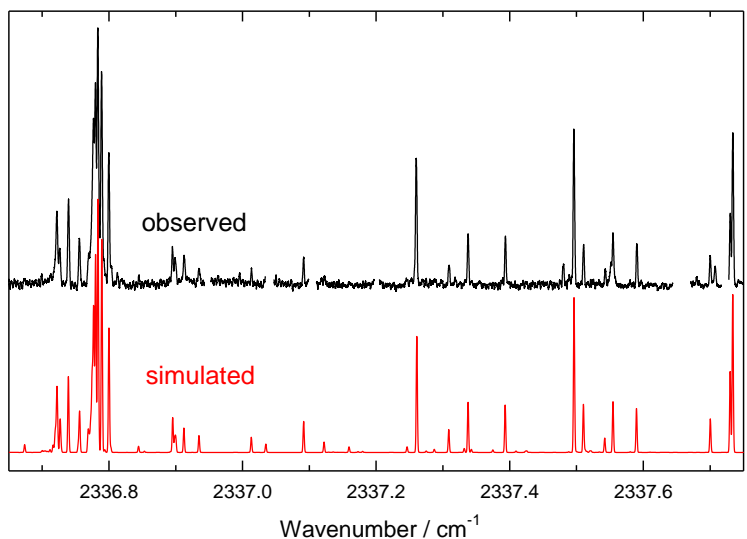

Fig. 2. Observed and simulated spectra of $CO_2$-Ne in the region of the $CO_2$ $(01^11) \leftarrow (01^10)$ hot band. Gaps in the observed spectrum correspond to regions of $CO_2$ monomer absorption. The simulated spectrum includes both the i-p and o-p modes, which are highly mixed by the Coriolis interaction.

### 2.2. Hot band, $CO_2$ $(01^11) \leftarrow (01^10)$

Part of the observed spectrum in the region of the $CO_2$ $(01^11) - (01^10)$ hot band is shown in Fig. 2. Observation of this weak spectrum is possible because a small fraction of $CO_2$ molecules remain "trapped" in the $(01^10)$ vibrational state following the supersonic jet expansion. The presence of the Ne atom breaks the symmetry of the doubly degenerate $CO_2$ $\nu_2$ bending mode into in-plane (i-p) and out-of-plane (o-p) components. As outlined previously for $CO_2$-Ar,[1] in the $C_{2v}$ point group these i-p and o-p modes have $A_1$ and $B_1$ symmetry, respectively, for the lower $(01^10)$ vibrational state, and $B_2$ and $A_2$ symmetry for the upper $(01^11)$ state. The hot band spectrum has $b$-type selection rules, the same as the fundamental. The i-p component is $B_2 \leftarrow A_1$ with $K_a$ = odd $\leftarrow$ even, and the o-p component is $A_2 \leftarrow B_1$ with $K_a$ = even $\leftarrow$ odd. There is strong $b$-type Coriolis mixing between the i-p and o-p modes, characterized by the matrix element

$$\langle \text{i-p}, J, k|H|\text{o-p}, J, k \pm 1\rangle = \tfrac{1}{2}\, \xi_b \times [J(J+1) - k(k\pm 1)]^{1/2},$$

where $k$ is signed $K_a$ and $\xi_b$ is the Coriolis parameter, related to the usual dimensionless zeta parameter and the $B$ rotational constant by $\xi_b = 2B\zeta_b$. In the present case, we expect $\zeta_b \approx 1$, as was found for $CO_2$-Ar.[1]

Table 2. Molecular parameters for the $(01^11) \leftarrow (01^10)$ hot band of $CO_2 – Ne$ (in cm$^{-1}$).[a]

|  | $(01^10)$ i-p | $(01^10)$ o-p | $(01^11)$ i-p | $(01^11)$ o-p |
|---|---|---|---|---|
| $\sigma_0$ | X [b] | 0.05659(42)+X | 2336.7749(3)+X | 2336.8240(5)+X |
| $A$ | 0.401628(100) | 0.400808(106) | 0.398302(77) | 0.397531(104) |
| $B$ | 0.115030(33) | 0.115500(51) | 0.114550(39) | 0.115332(28) |
| $C$ | 0.088046(38) | 0.087962(26) | 0.087695(34) | 0.088011(40) |
| $10^5 \times \Delta_K$ | [-6.39] | [-6.39] | [-6.39] | [-6.39] |
| $10^5 \times \Delta_{JK}$ | 5.29(40) | 7.29(62) | [5.29] | [7.29] |
| $10^6 \times \Delta_J$ | 5.80(44) | 4.36(41) | [5.80] | [4.36] |
| $10^5 \times \delta_K$ | [4.5224] | [4.5224] | [4.5224] | [4.5224] |
| $10^6 \times \delta_J$ | [5.2279] | [5.2279] | [5.2279] | [5.2279] |
| $\xi_b$ | 0.235640(60) |  | 0.235307(55) |  |

[a] Quantities in parentheses correspond to 1σ from the least-squares fit, in units of the last quoted digit. The ground and excited state centrifugal distortion parameters $\Delta_{JK}$ and $\Delta_J$ were constrained to be equal as indicated. Other centrifugal distortion parameters were fixed at ground state values (Table 1).

[b] X is equal to the free $CO_2$ $\nu_2$ frequency (667.380 cm$^{-1}$) plus or minus a (small) unknown vibrational shift.

The strongest feature in the hot band spectrum is a partly resolved $Q$-branch around 2336.78 cm$^{-1}$ (Fig. 2), which is accompanied by a weaker and more widely spaced $Q$-branch about 0.05 cm$^{-1}$ lower in wavenumber. The proximity of these $Q$-branches is a clue that the splitting between the i-p and o-p modes is quite small. Once we established the correct splitting by trial and error, it was possible to get a good simulation of the spectrum using PGOPHER. We assigned a total of 112 transitions of $CO_2$-$^{20}$Ne in the band and fitted them with an rms error of 0.00048 cm$^{-1}$ to obtain the parameters listed in Table 2. It was also possible to assign 26 transitions of $CO_2$-$^{22}$Ne, but these were not sufficient to obtain any really new information. Basically the $CO_2$-$^{22}$Ne spectrum could be very well reproduced by simply scaling the rotational constants from $^{20}$Ne to $^{22}$Ne using the fundamental band ratios (Table 1) and by assuming the same small change in the vibrational shift as for the fundamental. Line positions for both isotopologues are given as Supplementary Information.

The i-p to o-p splitting was determined to be 0.057 cm$^{-1}$ in the lower state, (01$^1$0), and 0.049 cm$^{-1}$ in the upper state, (01$^1$1), with o-p lying above i-p. These compare to previously reported splittings of 0.877 cm$^{-1}$ for $CO_2$-Ar,[1] 2.140 cm$^{-1}$ for $CO_2$-Xe,[2] and 2.307 cm$^{-1}$ for $CO_2$-$N_2$.[16] The values of $\xi_b$ in Table 2 correspond to $\zeta_b \approx 1.02$, depending slightly on what value is used for $B$, similar to what was observed for $CO_2$-Ar, -Xe, and -$N_2$. In the present case, the i-p to o-p splitting for $CO_2$-Ne is very small compared to the $K_a$ rotational level spacing, so Coriolis mixing is virtually complete. Complete mixing means that the quantum labels i-p or o-p, $K_a$, and $K_c$, are not very meaningful. For example, the transitions in the strong $Q$-branch at 2336.78 cm$^{-1}$ are best labeled as $K_a = 0$ (o-p) ← 0 (i-p), even though such transitions would be forbidden in the absence of Coriolis mixing. Similarly, the weaker $Q$-branch is nominally $K_a = 1$ (i-p) ← 1 (o-p).

The $CO_2$-Ne band origins from Table 2 represent vibrational shifts of +0.142 and +0.134 cm$^{-1}$ for the i-p and o-p modes, respectively, relative to the $(01^11) \leftarrow (01^10)$ hot band origin of the free $CO_2$ molecule, which is 2336.633 cm$^{-1}$.[17] These are very similar to the blue shift of +0.136 cm$^{-1}$ as determined above for the fundamental band. Although the hot band spectrum reveals the i-p to o-p splitting, it does not yield the actual $CO_2$ $v_2$ bending frequency for $CO_2$-Ne. This frequency is represented by X in Table 2, and it will probably be quite close ($< 0.5$ cm$^{-1}$?) to the free $CO_2$ value of 667.38 cm$^{-1}$. Our predicted $CO_2$-Ne spectrum in the $CO_2$ $v_2$ region for a temperature of 2 K is shown in Fig. 3. Here the i-p and o-p modes, which have *a*- and *c*-type selection rules, are shown separately. But, as mentioned above, they are highly mixed so these labels are only approximate. The strongest features in the predicted spectrum are *Q*-branches with $K_a = 0$ (i-p) $\leftarrow$ 0 and $K_a = 1$ (o-p) $\leftarrow$ 0 (the former would of course be forbidden in the absence of Coriolis mixing).

The observed $CO_2$ bending mode splittings in $CO_2$-Rg complexes are compared in Fig. 4 by plotting them as a function of Rg atomic polarizability, as is commonly done for vibrational shifts. There is an almost perfect linear relationship for Ne, Ar, and Kr, while the splitting observed for $CO_2$-Xe is a bit smaller than expected on the basis of linear extrapolation. A vertical line indicates the polarizability of He, and we see that extrapolation of the Ne-Ar-Kr line suggests that for $CO_2$-He we might expect a negative splitting (that is, with the o-p mode below the i-p mode). Unfortunately, we have so far been unable to detect the $CO_2$-He spectrum in the hot band region.

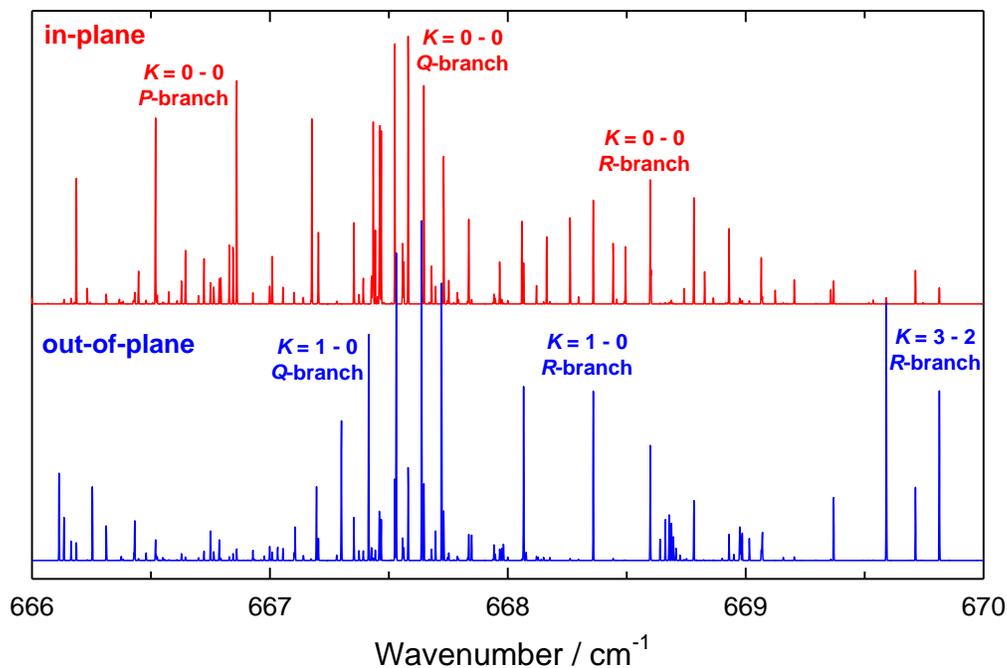

Fig. 3. Predicted spectrum of $CO_2$-Ne in the region of the $CO_2$ $\nu_2$ fundamental band for a temperature of 2K. Here it is assumed that the unknown vibrational shift relative to the free $CO_2$ molecule is zero. So the actual spectrum, not yet observed, will be shifted up or down from this simulation (probably by less than 0.5 cm$^{-1}$).

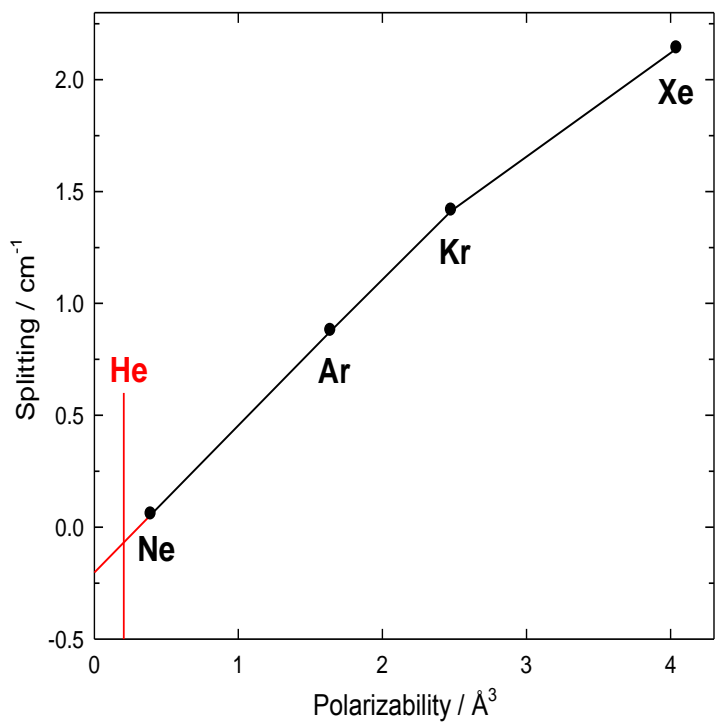

Fig. 4.  In-plane / out-of-plane mode splittings for $CO_2$-Rg complexes in the $\nu_2$ bending state of $CO_2$ as a function of atomic polarizability. Splitting values: Ne (present work), Ar[1], Kr[3] (to be published), and Xe.[2] Note the almost perfect linear relationship for Ne-Ar-Kr. The polarizability of He is indicated by the vertical red line, and extrapolation (in red) suggests that the splitting for $CO_2$-He (not yet determined) might be negative (i-p above o-p).

## 3. Conclusions

In conclusion, the infrared spectrum of the weakly-bound complex $CO_2$-Ne has been studied in the region of the $CO_2$ $\nu_3$ asymmetric stretch, using a tunable optical parametric oscillator source to probe a pulsed slit jet supersonic expansion. The fundamental band has been assigned for both $CO_2$-$^{20}$Ne and $CO_2$-$^{22}$Ne and fitted including available microwave data.6 The resulting molecular parameters (Table 1) should be the best currently available. In addition, the relatively weak hot band corresponding to the $CO_2$ $(01^11) - (01^10)$ transition has been detected for $CO_2$-Ne. Its analysis yields a measurement of the symmetry breaking of the degenerate $CO_2$ $\nu_2$ bend into in-plane and out-of-plane components, which turns out to have a magnitude of about 0.057 cm$^{-1}$ in the $(01^10)$ state, with o-p above i-p. This splitting is a sensitive probe of vibrational dynamics, and we have now published determinations for $CO_2$ with Ne, Ar and Xe.[1,2] In this context, $CO_2$-He[18,19] would be especially interesting: could the ordering of the i-p and o-p modes be reversed, as suggested by Fig. 4?

**Supplementary Information**

Supplementary Information includes tables giving observed and fitted line positions for $CO_2$-Ne.

**Acknowledgements**

The financial support of the Natural Sciences and Engineering Research Council of Canada is gratefully acknowledged.

Appendix to:

New infrared spectra of CO2-Ne:

Fundamental for CO2-22Ne isotopologue, intermolecular bend,

and symmetry breaking of the intramolecular CO2 bend

Table A-1. Observed transitions of the fundamental band of CO2-20Ne (units of 1/cm)

```
**********************************************************************
 J' Ka' Kc'   J" Ka" Kc" Observed    Calc     Obs-Calc Obs-Calc
**********************************************************************
```

| J' | Ka' | Kc' | J" | Ka" | Kc" | Observed | Calc | Obs-Calc | Obs-Calc |
|----|-----|-----|----|-----|-----|----------|------|----------|----------|
| 8  | 3   | 6   | 9  | 4   | 5   | 2345.3551 | 2345.3549 | 0.0003 | |
| 8  | 3   | 5   | 9  | 4   | 6   | 2345.3919 | 2345.3914 | 0.0004 | |
| 7  | 3   | 5   | 8  | 4   | 4   | 2345.5514 | 2345.5516 | -0.0002 | |
| 7  | 3   | 4   | 8  | 4   | 5   | 2345.5684 | 2345.5684 | 0.0001 | |
| 6  | 3   | 4   | 7  | 4   | 3   | 2345.7472 | 2345.7473 | -0.0001 | |
| 6  | 3   | 3   | 7  | 4   | 4   | 2345.7540 | 2345.7540 | -0.0000 | |
| 4  | 3   | 1   | 5  | 4   | 2   | 2346.1399 | 2346.1406 | | -0.0007 |
| 4  | 3   | 2   | 5  | 4   | 1   | 2346.1399 | 2346.1400 | | -0.0002 |
|    |     |     |    |     |     | Blend    | 2346.1403 | -0.0004 | |
| 3  | 3   | 0   | 4  | 4   | 1   | 2346.3382 | 2346.3379 | | 0.0002 |
| 3  | 3   | 1   | 4  | 4   | 0   | 2346.3382 | 2346.3379 | | 0.0003 |
|    |     |     |    |     |     | Blend    | 2346.3379 | 0.0003 | |
| 6  | 1   | 6   | 7  | 2   | 5   | 2346.4741 | 2346.4741 | -0.0000 | |
| 5  | 1   | 5   | 6  | 2   | 4   | 2346.8384 | 2346.8385 | -0.0001 | |
| 7  | 1   | 6   | 8  | 2   | 7   | 2347.1294 | 2347.1295 | -0.0000 | |

| | | | | | | | | |
|---|---|---|---|---|---|---|---|---|
| 4 | 3 | 1 | 4 | 4 | 0 | 2347.1454 | 2347.1456 | -0.0002 |
| 4 | 3 | 2 | 4 | 4 | 1 | 2347.1454 | 2347.1451 | 0.0004 |
| | | | | Blend | | 2347.1453 | 0.0001 | |
| 6 | 3 | 4 | 6 | 4 | 3 | 2347.1555 | 2347.1556 | -0.0002 |
| 6 | 1 | 5 | 7 | 2 | 6 | 2347.2430 | 2347.2429 | 0.0001 |
| 5 | 1 | 4 | 6 | 2 | 5 | 2347.3657 | 2347.3656 | 0.0001 |
| 3 | 1 | 3 | 4 | 2 | 2 | 2347.4549 | 2347.4549 | -0.0000 |
| 4 | 1 | 3 | 5 | 2 | 4 | 2347.4999 | 2347.4998 | 0.0001 |
| 8 | 1 | 8 | 9 | 0 | 9 | 2347.6324 | 2347.6326 | -0.0002 |
| 3 | 1 | 2 | 4 | 2 | 3 | 2347.6467 | 2347.6468 | -0.0001 |
| 2 | 1 | 2 | 3 | 2 | 1 | 2347.7157 | 2347.7158 | -0.0001 |
| 2 | 1 | 1 | 3 | 2 | 2 | 2347.8075 | 2347.8075 | -0.0000 |
| 7 | 1 | 7 | 8 | 0 | 8 | 2347.8316 | 2347.8316 | 0.0000 |
| 1 | 1 | 1 | 2 | 2 | 0 | 2347.9529 | 2347.9529 | -0.0000 |
| 1 | 1 | 0 | 2 | 2 | 1 | 2347.9823 | 2347.9823 | -0.0000 |
| 6 | 1 | 6 | 7 | 0 | 7 | 2348.0388 | 2348.0386 | 0.0002 |
| 6 | 1 | 6 | 6 | 2 | 5 | 2348.0736 | 2348.0735 | 0.0001 |
| 5 | 1 | 5 | 5 | 2 | 4 | 2348.1604 | 2348.1604 | -0.0000 |
| 4 | 1 | 4 | 4 | 2 | 3 | 2348.2326 | 2348.2324 | 0.0001 |
| 5 | 1 | 5 | 6 | 0 | 6 | 2348.2544 | 2348.2543 | 0.0001 |
| 3 | 1 | 3 | 3 | 2 | 2 | 2348.2895 | 2348.2896 | -0.0001 |
| 2 | 1 | 2 | 2 | 2 | 1 | 2348.3319 | 2348.3320 | -0.0002 |
| 2 | 1 | 1 | 2 | 2 | 0 | 2348.4126 | 2348.4127 | -0.0001 |
| 3 | 1 | 2 | 3 | 2 | 1 | 2348.4451 | 2348.4453 | -0.0002 |
| 4 | 1 | 4 | 5 | 0 | 5 | 2348.4784 | 2348.4777 | 0.0007 |
| 4 | 1 | 3 | 4 | 2 | 2 | 2348.4784 | 2348.4794 | -0.0010 |
| | | | | Blend | | 2348.4784 | -0.0001 | |
| 5 | 1 | 4 | 5 | 2 | 3 | 2348.5079 | 2348.5080 | -0.0001 |
| 7 | 1 | 6 | 7 | 2 | 5 | 2348.5201 | 2348.5202 | -0.0000 |
| 6 | 1 | 5 | 6 | 2 | 4 | 2348.5237 | 2348.5237 | -0.0001 |
| 3 | 1 | 3 | 4 | 0 | 4 | 2348.7060 | 2348.7059 | 0.0001 |
| 3 | 1 | 3 | 2 | 2 | 0 | 2348.8947 | 2348.8948 | -0.0001 |
| 2 | 1 | 2 | 3 | 0 | 3 | 2348.9347 | 2348.9346 | 0.0001 |
| 3 | 1 | 2 | 2 | 2 | 1 | 2349.0613 | 2349.0615 | -0.0002 |
| 5 | 1 | 5 | 4 | 2 | 2 | 2349.1400 | 2349.1400 | 0.0000 |
| 1 | 1 | 1 | 2 | 0 | 2 | 2349.1589 | 2349.1589 | 0.0000 |
| 6 | 1 | 6 | 5 | 2 | 3 | 2349.2157 | 2349.2158 | -0.0002 |
| 4 | 1 | 3 | 3 | 2 | 2 | 2349.3141 | 2349.3141 | 0.0000 |
| 1 | 1 | 0 | 1 | 0 | 1 | 2349.5907 | 2349.5906 | 0.0001 |
| 2 | 1 | 1 | 2 | 0 | 2 | 2349.6188 | 2349.6187 | 0.0001 |
| 3 | 1 | 2 | 3 | 0 | 3 | 2349.6642 | 2349.6641 | 0.0001 |
| 4 | 1 | 3 | 4 | 0 | 4 | 2349.7306 | 2349.7304 | 0.0002 |
| 1 | 1 | 1 | 0 | 0 | 0 | 2349.7662 | 2349.7662 | -0.0000 |
| 5 | 1 | 4 | 5 | 0 | 5 | 2349.8217 | 2349.8214 | 0.0003 |
| 6 | 1 | 5 | 5 | 2 | 4 | 2349.8458 | 2349.8456 | 0.0001 |
| 2 | 1 | 2 | 1 | 0 | 1 | 2349.9403 | 2349.9403 | -0.0000 |
| 3 | 1 | 3 | 2 | 0 | 2 | 2350.1008 | 2350.1007 | 0.0001 |
| 4 | 1 | 4 | 3 | 0 | 3 | 2350.2499 | 2350.2497 | 0.0002 |
| 5 | 1 | 5 | 4 | 0 | 4 | 2350.3910 | 2350.3911 | -0.0001 |
| 6 | 1 | 6 | 5 | 0 | 5 | 2350.5293 | 2350.5293 | 0.0001 |
| 6 | 3 | 3 | 6 | 2 | 4 | 2350.6547 | 2350.6548 | -0.0001 |
| 7 | 1 | 7 | 6 | 0 | 6 | 2350.6684 | 2350.6684 | -0.0000 |
| 5 | 3 | 2 | 5 | 2 | 3 | 2350.6987 | 2350.6987 | -0.0000 |
| 4 | 3 | 1 | 4 | 2 | 2 | 2350.7284 | 2350.7285 | -0.0001 |
| 3 | 3 | 0 | 3 | 2 | 1 | 2350.7464 | 2350.7464 | 0.0000 |
| 8 | 1 | 8 | 7 | 0 | 7 | 2350.8114 | 2350.8113 | 0.0001 |
| 9 | 1 | 9 | 8 | 0 | 8 | 2350.9591 | 2350.9590 | 0.0001 |
| 3 | 3 | 0 | 2 | 2 | 1 | 2351.3615 | 2351.3626 | -0.0010 |
| 3 | 3 | 1 | 2 | 2 | 0 | 2351.3615 | 2351.3606 | 0.0009 |
| | | | | | Blend | | 2351.3616 | -0.0001 |
| 4 | 3 | 2 | 3 | 2 | 1 | 2351.5535 | 2351.5535 | 0.0000 |
| 4 | 3 | 1 | 3 | 2 | 2 | 2351.5631 | 2351.5632 | -0.0001 |
| 5 | 3 | 3 | 4 | 2 | 2 | 2351.7378 | 2351.7377 | 0.0002 |
| 5 | 3 | 2 | 4 | 2 | 3 | 2351.7669 | 2351.7669 | 0.0001 |
| 6 | 3 | 4 | 5 | 2 | 3 | 2351.9088 | 2351.9088 | -0.0000 |
| 6 | 3 | 3 | 5 | 2 | 4 | 2351.9766 | 2351.9767 | -0.0001 |
| 7 | 3 | 4 | 6 | 2 | 5 | 2352.1968 | 2352.1967 | 0.0001 |
| 8 | 3 | 6 | 7 | 2 | 5 | 2352.1968 | 2352.1981 | -0.0013 |
| | | | | | Blend | | 2352.1971 | -0.0002 |
| 8 | 3 | 5 | 7 | 2 | 6 | 2352.4317 | 2352.4322 | -0.0005 |
| 5 | 5 | 0 | 4 | 4 | 1 | 2352.9136 | 2352.9134 | 0.0002 |
| 5 | 5 | 1 | 4 | 4 | 0 | 2352.9136 | 2352.9134 | 0.0002 |
|   |   |   |   |   |   | Blend | 2352.9134 | 0.0002 |
| 6 | 5 | 1 | 5 | 4 | 2 | 2353.1009 | 2353.1014 | -0.0004 |
| 6 | 5 | 2 | 5 | 4 | 1 | 2353.1009 | 2353.1014 | -0.0004 |
|   |   |   |   |   |   | Blend | 2353.1014 | -0.0004 |
| 6 | 3 | 4 | 5 | 0 | 5 | 2353.2226 | 2353.2223 | 0.0004 |
| 7 | 5 | 2 | 6 | 4 | 3 | 2353.2870 | 2353.2866 | 0.0005 |
| 7 | 5 | 3 | 6 | 4 | 2 | 2353.2870 | 2353.2865 | 0.0006 |
|   |   |   |   |   |   | Blend | 2353.2865 | 0.0005 |
| 7 | 3 | 5 | 6 | 0 | 6 | 2353.4784 | 2353.4788 | -0.0004 |

*******************************************************************

Table A-2. Observed transitions in the microwave spectrum of CO2-20Ne

MW data from: Y. Xu and W. Jäger, J. Mol. Spectrosc. 192, 435-440 (1998).

************************************************************************

| J' Ka' Kc' | J" Ka" Kc" | Obs(MHz) | Calc(MHz) | Obs-Calc(MHz) |
|---|---|---|---|---|
| 1 0 1 | 0 0 0 | 6088.2440 | 6088.2413 | 0.0027 |
| 2 0 2 | 1 0 1 | 12117.9430 | 12117.9444 | -0.0014 |
| 3 0 3 | 2 0 2 | 18032.3303 | 18032.3309 | -0.0006 |
| 4 0 4 | 3 0 3 | 23782.8865 | 23782.8861 | 0.0004 |
| 3 2 2 | 2 2 1 | 18198.6807 | 18198.6791 | 0.0016 |
| 3 2 1 | 2 2 0 | 18418.0132 | 18418.0146 | -0.0014 |
| 4 2 3 | 3 2 2 | 24211.8532 | 24211.8539 | -0.0007 |
| 4 2 2 | 3 2 1 | 24749.1378 | 24749.1372 | 0.0006 |

************************************************************************

Table A-3. Observed transitions of the fundamental band of CO2-22Ne (units of 1/cm)

```
********************************************************************

 J' Ka' Kc'   J" Ka" Kc"  Observed    Calc     Obs-Calc  Obs-Calc

********************************************************************

 4  3  1     5  4  2   2346.1594   2346.1598    -0.0003

 4  3  2     5  4  1   2346.1594   2346.1594     0.0000

                 Blend   2346.1596  -0.0002

 3  3  0     4  4  1   2346.3469   2346.3466     0.0003

 3  3  1     4  4  0   2346.3469   2346.3465     0.0003

                 Blend   2346.3466   0.0003

 4  1  4     5  2  3   2347.2324   2347.2325   -0.0001

 6  1  5     7  2  6   2347.2772   2347.2771    0.0001

 5  1  4     6  2  5   2347.3967   2347.3966    0.0001

 4  1  3     5  2  4   2347.5269   2347.5267    0.0002

 3  1  2     4  2  3   2347.6684   2347.6684   -0.0001

 2  1  2     3  2  1   2347.7413   2347.7414   -0.0001

 2  1  1     3  2  2   2347.8225   2347.8225    0.0000

 1  1  1     2  2  0   2347.9630   2347.9630    0.0000

 1  1  0     2  2  1   2347.9891   2347.9891    0.0000

 6  1  6     6  2  5   2348.0915   2348.0915    0.0000

 6  1  6     7  0  7   2348.1212   2348.1212    0.0000

 5  1  5     5  2  4   2348.1689   2348.1691   -0.0002

 2  1  2     2  2  1   2348.3221   2348.3223   -0.0001

 2  1  1     2  2  0   2348.3948   2348.3947    0.0001
```

| | | | | | | | | |
|---|---|---|---|---|---|---|---|---|
| 3 | 1 | 2 | 3 | 2 | 1 | 2348.4245 | 2348.4247 | -0.0002 |
| 4 | 1 | 3 | 4 | 2 | 2 | 2348.4573 | 2348.4573 | -0.0000 |
| 5 | 1 | 4 | 5 | 2 | 3 | 2348.4866 | 2348.4868 | -0.0002 |
| 4 | 1 | 4 | 5 | 0 | 5 | 2348.5441 | 2348.5441 | -0.0000 |
| 3 | 1 | 3 | 4 | 0 | 4 | 2348.7617 | 2348.7616 | 0.0001 |
| 2 | 1 | 2 | 3 | 0 | 3 | 2348.9783 | 2348.9782 | 0.0000 |
| 1 | 1 | 1 | 2 | 0 | 2 | 2349.1899 | 2349.1899 | 0.0000 |
| 1 | 1 | 0 | 1 | 0 | 1 | 2349.5968 | 2349.5967 | 0.0000 |
| 4 | 1 | 3 | 4 | 0 | 4 | 2349.7195 | 2349.7195 | 0.0001 |
| 5 | 1 | 4 | 5 | 0 | 5 | 2349.7986 | 2349.7984 | 0.0001 |
| 3 | 1 | 3 | 2 | 0 | 2 | 2350.0838 | 2350.0835 | 0.0003 |
| 5 | 1 | 5 | 4 | 0 | 4 | 2350.3619 | 2350.3619 | -0.0000 |
| 7 | 1 | 7 | 6 | 0 | 6 | 2350.6245 | 2350.6244 | 0.0001 |
| 4 | 3 | 2 | 4 | 2 | 3 | 2350.7837 | 2350.7835 | 0.0001 |
| 3 | 3 | 1 | 3 | 2 | 2 | 2350.7837 | 2350.7843 | -0.0007 |
| | | | | | Blend | | 2350.7839 | -0.0003 |
| 4 | 3 | 2 | 3 | 2 | 1 | 2351.5395 | 2351.5398 | -0.0003 |
| 5 | 5 | 0 | 4 | 4 | 1 | 2352.9090 | 2352.9087 | 0.0003 |
| 5 | 5 | 1 | 4 | 4 | 0 | 2352.9090 | 2352.9087 | 0.0003 |
| | | | | | Blend | | 2352.9087 | 0.0003 |
| 6 | 5 | 1 | 5 | 4 | 2 | 2353.0871 | 2353.0872 | -0.0001 |
| 6 | 5 | 2 | 5 | 4 | 1 | 2353.0871 | 2353.0872 | -0.0001 |
| | | | | | Blend | | 2353.0872 | -0.0001 |
| 7 | 5 | 2 | 6 | 4 | 3 | 2353.2632 | 2353.2634 | -0.0001 |

```
 7  5  3    6  4  2  2353.2632        2353.2633       -0.0001
                        Blend         2353.2633 -0.0001
 3  1  3    4  0  4  2348.7617        2348.7616  0.0001
```

******************************************************************

Table A-4. Observed transitions in the microwave spectrum of CO2-22Ne

MW data from: Y. Xu and W. Jäger, J. Mol. Spectrosc. 192, 435-440 (1998).

**********************************************************************

 J' Ka' Kc'   J" Ka" Kc" Obs(MHz)    Calc(MHz)   Obs-Calc(MHz)

**********************************************************************

 1 0 1     0 0 0  5752.5423   5752.5425  -0.0002

 2 0 2     1 0 1 11458.7128  11458.7126   0.0002

 3 0 3     2 0 2 17073.2426  17073.2428  -0.0002

 3 2 2     2 2 1 17198.8324  17198.8323   0.0001

 3 2 1     2 2 0 17371.6715  17371.6716  -0.0001

 4 0 4     3 0 3 22556.0274  22556.0273   0.0001

 4 2 3     3 2 2 22888.5325  22888.5325  -0.0000

 4 2 2     3 2 1 23313.7122  23313.7122   0.0000

**********************************************************************

Table A-5. Observed transitions of the combination band of CO2-20Ne (units of 1/cm)

**************************************************************

| J' | Ka' | Kc' | J" | Ka" | Kc" | Observed | Calc | Obs-Calc |
|---|---|---|---|---|---|---|---|---|
| 6 | 2 | 4 | 7 | 2 | 5 | 2365.3207 | 2365.3190 | 0.0017 |
| 6 | 0 | 6 | 7 | 0 | 7 | 2365.3716 | 2365.3707 | 0.0009 |
| 5 | 0 | 5 | 6 | 0 | 6 | 2365.6291 | 2365.6290 | 0.0000 |
| 5 | 2 | 3 | 6 | 2 | 4 | 2365.6546 | 2365.6547 | -0.0002 |
| 5 | 2 | 4 | 6 | 2 | 5 | 2365.7207 | 2365.7200 | 0.0007 |
| 4 | 0 | 4 | 5 | 0 | 5 | 2365.8779 | 2365.8770 | 0.0009 |
| 4 | 2 | 2 | 5 | 2 | 3 | 2365.9698 | 2365.9702 | -0.0004 |
| 4 | 2 | 3 | 5 | 2 | 4 | 2366.0086 | 2366.0093 | -0.0007 |
| 3 | 0 | 3 | 4 | 0 | 4 | 2366.1183 | 2366.1177 | 0.0006 |
| 3 | 2 | 1 | 4 | 2 | 2 | 2366.2622 | 2366.2624 | -0.0002 |
| 3 | 2 | 2 | 4 | 2 | 3 | 2366.2817 | 2366.2820 | -0.0003 |
| 2 | 0 | 2 | 3 | 0 | 3 | 2366.3517 | 2366.3515 | 0.0002 |
| 2 | 2 | 0 | 3 | 2 | 1 | 2366.5305 | 2366.5302 | 0.0003 |
| 2 | 2 | 1 | 3 | 2 | 2 | 2366.5381 | 2366.5379 | 0.0002 |
| 1 | 0 | 1 | 2 | 0 | 2 | 2366.5764 | 2366.5773 | -0.0009 |
| 0 | 0 | 0 | 1 | 0 | 1 | 2366.7915 | 2366.7929 | -0.0014 |
| 5 | 2 | 4 | 5 | 2 | 3 | 2366.8627 | 2366.8624 | 0.0002 |
| 5 | 2 | 3 | 5 | 2 | 4 | 2366.9764 | 2366.9766 | -0.0002 |
| 4 | 2 | 3 | 4 | 2 | 2 | 2366.9883 | 2366.9889 | -0.0006 |
| 4 | 2 | 2 | 4 | 2 | 3 | 2367.0381 | 2367.0385 | -0.0004 |
| 3 | 2 | 2 | 3 | 2 | 1 | 2367.0801 | 2367.0805 | -0.0004 |
| 3 | 2 | 1 | 3 | 2 | 2 | 2367.0969 | 2367.0971 | -0.0001 |
| 2 | 2 | 1 | 2 | 2 | 0 | 2367.1438 | 2367.1431 | 0.0007 |
| 2 | 2 | 0 | 2 | 2 | 1 | 2367.1469 | 2367.1464 | 0.0005 |
| 1 | 0 | 1 | 0 | 0 | 0 | 2367.1842 | 2367.1846 | -0.0004 |
| 2 | 0 | 2 | 1 | 0 | 1 | 2367.3575 | 2367.3573 | 0.0002 |
| 3 | 0 | 3 | 2 | 0 | 2 | 2367.5134 | 2367.5125 | 0.0009 |
| 4 | 0 | 4 | 3 | 0 | 3 | 2367.6503 | 2367.6491 | 0.0011 |
| 3 | 2 | 2 | 2 | 2 | 1 | 2367.6971 | 2367.6967 | 0.0004 |
| 3 | 2 | 1 | 2 | 2 | 0 | 2367.7029 | 2367.7023 | 0.0006 |
| 5 | 0 | 5 | 4 | 0 | 4 | 2367.7669 | 2367.7660 | 0.0009 |
| 4 | 2 | 3 | 3 | 2 | 2 | 2367.8243 | 2367.8236 | 0.0006 |
| 4 | 2 | 2 | 3 | 2 | 1 | 2367.8374 | 2367.8369 | 0.0005 |
| 6 | 0 | 6 | 5 | 0 | 5 | 2367.8601 | 2367.8616 | -0.0015 |
| 5 | 2 | 4 | 4 | 2 | 3 | 2367.9293 | 2367.9307 | -0.0014 |
| 7 | 0 | 7 | 6 | 0 | 6 | 2367.9318 | 2367.9345 | -0.0027 |
| 5 | 2 | 3 | 4 | 2 | 2 | 2367.9544 | 2367.9563 | -0.0019 |
| 8 | 0 | 8 | 7 | 0 | 7 | 2367.9843 | 2367.9830 | 0.0012 |
| 6 | 2 | 5 | 5 | 2 | 4 | 2368.0190 | 2368.0171 | 0.0018 |
| 6 | 2 | 4 | 5 | 2 | 3 | 2368.0596 | 2368.0608 | -0.0012 |

************************************************************

Table A-6. Observed transitions of the combination band of CO2-22Ne (units of 1/cm)

***********************************************************

| J' | Ka' | Kc' | J" | Ka" | Kc" | Observed | Calc | Obs-Calc |
|---|---|---|---|---|---|---|---|---|
| 5 | 0 | 5 | 6 | 0 | 6 | 2365.7069 | 2365.7079 | -0.0010 |
| 5 | 2 | 4 | 6 | 2 | 5 | 2365.8065 | 2365.8062 | 0.0003 |
| 4 | 0 | 4 | 5 | 0 | 5 | 2365.9455 | 2365.9458 | -0.0003 |
| 4 | 2 | 2 | 5 | 2 | 3 | 2366.0495 | 2366.0493 | 0.0002 |
| 4 | 2 | 3 | 5 | 2 | 4 | 2366.0805 | 2366.0810 | -0.0005 |
| 3 | 0 | 3 | 4 | 0 | 4 | 2366.1760 | 2366.1760 | 0.0000 |
| 3 | 2 | 1 | 4 | 2 | 2 | 2366.3224 | 2366.3232 | -0.0007 |
| 3 | 2 | 2 | 4 | 2 | 3 | 2366.3384 | 2366.3389 | -0.0005 |
| 2 | 0 | 2 | 3 | 0 | 3 | 2366.3988 | 2366.3986 | 0.0001 |
| 1 | 0 | 1 | 2 | 0 | 2 | 2366.6124 | 2366.6126 | -0.0003 |
| 0 | 0 | 0 | 1 | 0 | 1 | 2366.8163 | 2366.8164 | -0.0002 |
| 4 | 2 | 3 | 4 | 2 | 2 | 2367.0111 | 2367.0116 | -0.0005 |
| 4 | 2 | 2 | 4 | 2 | 3 | 2367.0500 | 2367.0502 | -0.0003 |
| 3 | 2 | 1 | 3 | 2 | 2 | 2367.1078 | 2367.1081 | -0.0002 |
| 2 | 2 | 0 | 2 | 2 | 1 | 2367.1540 | 2367.1550 | -0.0010 |
| 2 | 2 | 1 | 2 | 2 | 0 | 2367.1540 | 2367.1524 | 0.0016 |
| 2 | 0 | 2 | 1 | 0 | 1 | 2367.3505 | 2367.3504 | 0.0002 |
| 3 | 0 | 3 | 2 | 0 | 2 | 2367.4982 | 2367.4979 | 0.0003 |
| 4 | 0 | 4 | 3 | 0 | 3 | 2367.6286 | 2367.6282 | 0.0004 |
| 3 | 2 | 2 | 2 | 2 | 1 | 2367.6762 | 2367.6760 | 0.0002 |

```
3  2  1     2  2  0 2367.6808 2367.6803  0.0005

5  0  5     4  0  4 2367.7409 2367.7401  0.0008

4  2  3     3  2  2 2367.7973 2367.7965  0.0008
```

***********************************************************

Table A-7. Observed transitions of the hot band of CO2-20Ne (units of 1/cm)

*********************************************************************

J' Ka' Kc'   J" Ka" Kc" Observed   Calc   Obs-Calc Obs-Calc

*********************************************************************

 3  2  2    4  3  1 2334.4230  2334.4230   0.0001

 3  2  1    4  3  2 2334.5275  2334.5279  -0.0004

 2  2  1    3  3  0 2334.6462  2334.6457   0.0005

 2  2  0    3  3  1 2334.6697  2334.6695   0.0002

 5  1  5    6  2  4 2334.9574  2334.9576  -0.0002

 2  1  2    3  2  1 2334.9886  2334.9877   0.0009

 6  0  6    7  1  7 2335.0148  2335.0138   0.0010

 4  0  4    5  2  4 2335.0991  2335.1001  -0.0010

 5  0  5    6  1  6 2335.1749  2335.1746   0.0003

 4  2  3    4  3  2 2335.2316  2335.2319  -0.0004

 3  2  2    3  3  1 2335.2618  2335.2616   0.0002

 3  1  2    4  2  3 2335.2951  2335.2949   0.0002

 5  1  5    5  2  4 2335.3339  2335.3348  -0.0009

 4  1  3    5  1  5 2335.3537  2335.3541  -0.0004

 1  1  1    2  2  0 2335.3673  2335.3675  -0.0001

 2  1  1    3  2  2 2335.4476  2335.4470   0.0006

 4  1  4    4  2  3 2335.4604  2335.4601   0.0003

 3  1  3    4  2  2 2335.4712  2335.4714  -0.0002

 3  0  3    4  1  4 2335.5534  2335.5542  -0.0008

 1  1  0    2  2  1 2335.5581  2335.5584  -0.0003

| | | | | | | | | |
|---|---|---|---|---|---|---|---|---|
| 5 | 0 | 5 | 5 | 2 | 3 | 2335.5771 | 2335.5769 | 0.0002 |
| 3 | 1 | 3 | 3 | 2 | 2 | 2335.5916 | 2335.5910 | 0.0006 |
| 4 | 1 | 3 | 4 | 2 | 2 | 2335.6450 | 2335.6451 | -0.0001 |
| 2 | 1 | 2 | 2 | 2 | 1 | 2335.7166 | 2335.7168 | -0.0002 |
| 2 | 1 | 2 | 3 | 1 | 2 | 2335.7585 | 2335.7587 | -0.0001 |
| 3 | 0 | 3 | 3 | 1 | 2 | 2335.7655 | 2335.7661 | -0.0006 |
| 2 | 0 | 2 | 3 | 1 | 3 | 2335.7763 | 2335.7770 | -0.0006 |
| 2 | 0 | 2 | 2 | 1 | 1 | 2335.9547 | 2335.9550 | -0.0004 |
| 3 | 1 | 2 | 3 | 2 | 1 | 2335.9997 | 2335.9985 | 0.0012 |
| 1 | 0 | 1 | 2 | 1 | 2 | 2336.0224 | 2336.0229 | -0.0004 |
| 2 | 1 | 1 | 2 | 2 | 0 | 2336.0448 | 2336.0449 | -0.0001 |
| 1 | 1 | 1 | 2 | 1 | 1 | 2336.0779 | 2336.0783 | -0.0004 |
| 5 | 1 | 5 | 6 | 0 | 6 | 2336.1874 | 2336.1874 | -0.0001 |
| 1 | 0 | 1 | 1 | 1 | 0 | 2336.2012 | 2336.2012 | -0.0001 |
| 0 | 0 | 0 | 1 | 1 | 1 | 2336.2789 | 2336.2786 | 0.0002 |
| 2 | 1 | 1 | 3 | 1 | 3 | 2336.5778 | 2336.5777 | 0.0000 |
| 2 | 0 | 2 | 1 | 1 | 1 | 2336.5924 | 2336.5923 | 0.0001 |
| 4 | 1 | 4 | 4 | 1 | 4 | 2336.7220 | 2336.7228 | -0.0008 |
| 3 | 1 | 3 | 3 | 1 | 3 | 2336.7220 | 2336.7217 | 0.0003 |
| | | | | Blend | | | 2336.7223 | -0.0003 |
| 3 | 0 | 3 | 2 | 1 | 2 | 2336.7263 | 2336.7265 | -0.0002 |
| 1 | 1 | 0 | 1 | 1 | 0 | 2336.7395 | 2336.7390 | 0.0004 |
| 2 | 1 | 1 | 2 | 1 | 1 | 2336.7557 | 2336.7558 | 0.0000 |
| 5 | 0 | 5 | 5 | 0 | 5 | 2336.7768 | 2336.7766 | 0.0002 |
| 4 | 1 | 3 | 4 | 0 | 4 | 2336.7798 | 2336.7794 | 0.0003 |
| 3 | 0 | 3 | 3 | 0 | 3 | 2336.7833 | 2336.7833 | 0.0001 |
| 2 | 0 | 2 | 2 | 0 | 2 | 2336.7890 | 2336.7891 | -0.0002 |
| 1 | 0 | 1 | 1 | 0 | 1 | 2336.7998 | 2336.8002 | -0.0004 |
| 4 | 1 | 3 | 3 | 1 | 3 | 2336.8951 | 2336.8954 | -0.0003 |
| 1 | 1 | 1 | 2 | 0 | 2 | 2336.9126 | 2336.9124 | 0.0002 |
| 5 | 1 | 5 | 4 | 2 | 2 | 2337.0134 | 2337.0133 | 0.0001 |
| 5 | 0 | 5 | 4 | 1 | 4 | 2337.0915 | 2337.0917 | -0.0003 |
| 1 | 1 | 1 | 0 | 0 | 0 | 2337.2601 | 2337.2609 | -0.0009 |
| 6 | 0 | 6 | 5 | 1 | 5 | 2337.3092 | 2337.3093 | -0.0001 |
| 1 | 1 | 0 | 1 | 0 | 1 | 2337.3377 | 2337.3380 | -0.0003 |
| 2 | 1 | 1 | 1 | 1 | 1 | 2337.3935 | 2337.3930 | 0.0005 |
| 2 | 1 | 2 | 1 | 0 | 1 | 2337.4961 | 2337.4964 | -0.0003 |
| 2 | 2 | 0 | 2 | 1 | 1 | 2337.5106 | 2337.5101 | 0.0005 |
| 7 | 0 | 7 | 6 | 1 | 6 | 2337.5431 | 2337.5429 | 0.0002 |
| 3 | 2 | 1 | 3 | 1 | 2 | 2337.5546 | 2337.5550 | -0.0004 |
| 2 | 1 | 1 | 2 | 0 | 2 | 2337.5904 | 2337.5899 | 0.0005 |
| 4 | 2 | 2 | 4 | 2 | 2 | 2337.7002 | 2337.7008 | -0.0006 |
| 3 | 1 | 2 | 2 | 1 | 2 | 2337.7305 | 2337.7299 | 0.0006 |
| 3 | 1 | 3 | 2 | 0 | 2 | 2337.7340 | 2337.7338 | 0.0002 |
| 3 | 1 | 2 | 3 | 0 | 3 | 2337.7876 | 2337.7867 | 0.0010 |
| 2 | 2 | 1 | 2 | 1 | 2 | 2337.8051 | 2337.8046 | 0.0006 |
| 5 | 1 | 4 | 5 | 2 | 3 | 2337.9028 | 2337.9032 | -0.0004 |
| 4 | 0 | 4 | 4 | 0 | 4 | 2337.9137 | 2337.9129 | 0.0008 |
| 3 | 2 | 2 | 3 | 1 | 3 | 2337.9240 | 2337.9242 | -0.0002 |
| 4 | 1 | 4 | 3 | 0 | 3 | 2337.9523 | 2337.9519 | 0.0004 |
| 2 | 2 | 1 | 1 | 1 | 0 | 2337.9836 | 2337.9829 | 0.0007 |
| 4 | 2 | 2 | 3 | 1 | 3 | 2338.0287 | 2338.0289 | -0.0002 |
| 4 | 2 | 3 | 4 | 1 | 4 | 2338.0465 | 2338.0472 | -0.0007 |
| 3 | 2 | 2 | 2 | 1 | 1 | 2338.1022 | 2338.1023 | 0.0000 |
| 3 | 3 | 0 | 3 | 2 | 1 | 2338.1282 | 2338.1275 | 0.0007 |
| 2 | 2 | 0 | 1 | 1 | 1 | 2338.1481 | 2338.1474 | 0.0007 |
| 5 | 1 | 5 | 4 | 0 | 4 | 2338.1481 | 2338.1476 | 0.0004 |
| | | | Blend | | | 2338.1475 | 0.0006 | |
| 5 | 1 | 4 | 4 | 2 | 3 | 2338.1549 | 2338.1553 | -0.0004 |
| 5 | 2 | 4 | 5 | 1 | 5 | 2338.1625 | 2338.1632 | -0.0007 |
| 4 | 3 | 2 | 4 | 2 | 3 | 2338.2550 | 2338.2552 | -0.0002 |
| 4 | 2 | 3 | 3 | 1 | 2 | 2338.2588 | 2338.2590 | -0.0003 |
| 5 | 2 | 3 | 4 | 1 | 4 | 2338.2970 | 2338.2974 | -0.0004 |
| 6 | 1 | 6 | 5 | 0 | 5 | 2338.3224 | 2338.3220 | 0.0003 |
| 5 | 2 | 4 | 4 | 2 | 2 | 2338.4538 | 2338.4543 | -0.0005 |
| 3 | 2 | 1 | 2 | 1 | 2 | 2338.5152 | 2338.5154 | -0.0002 |
| 6 | 2 | 4 | 5 | 1 | 5 | 2338.5468 | 2338.5477 | -0.0009 |
| 8 | 1 | 8 | 7 | 0 | 7 | 2338.6134 | 2338.6143 | -0.0009 |
| 6 | 2 | 5 | 5 | 2 | 3 | 2338.6701 | 2338.6702 | -0.0001 |
| 7 | 2 | 5 | 6 | 1 | 6 | 2338.7890 | 2338.7893 | -0.0003 |
| 3 | 3 | 1 | 2 | 2 | 0 | 2338.8297 | 2338.8296 | 0.0001 |
| 3 | 3 | 0 | 2 | 2 | 1 | 2338.8563 | 2338.8566 | -0.0003 |
| 7 | 2 | 6 | 6 | 2 | 4 | 2338.8919 | 2338.8918 | 0.0001 |
| 4 | 3 | 2 | 3 | 2 | 1 | 2338.9591 | 2338.9588 | 0.0004 |
| 8 | 2 | 6 | 7 | 1 | 7 | 2339.0307 | 2339.0298 | 0.0009 |
| 5 | 3 | 3 | 4 | 1 | 3 | 2339.0511 | 2339.0508 | 0.0003 |
| 4 | 3 | 1 | 3 | 2 | 2 | 2339.0736 | 2339.0741 | -0.0005 |
| 4 | 1 | 4 | 5 | 2 | 3 | 2335.2076 | 2335.2080 | -0.0004 |
| 5 | 2 | 3 | 5 | 1 | 4 | 2335.6343 | 2335.6341 | 0.0003 |
| 2 | 2 | 1 | 3 | 1 | 2 | 2336.8448 | 2336.8442 | 0.0006 |
| 3 | 1 | 3 | 2 | 1 | 1 | 2336.8992 | 2336.8998 | -0.0006 |
| 4 | 0 | 4 | 3 | 2 | 2 | 2336.8992 | 2336.8982 | 0.0010 |
| 2 | 1 | 2 | 1 | 1 | 0 | 2336.8992 | 2336.8974 | 0.0018 |
| | | | Blend | | | 2336.8988 | 0.0004 | |
| 4 | 1 | 4 | 3 | 1 | 2 | 2336.9349 | 2336.9347 | 0.0002 |
| 6 | 1 | 6 | 5 | 2 | 3 | 2337.1227 | 2337.1223 | 0.0004 |
| 4 | 2 | 2 | 3 | 2 | 2 | 2337.8202 | 2337.8204 | -0.0002 |
| 5 | 3 | 2 | 5 | 1 | 4 | 2337.9280 | 2337.9288 | -0.0008 |
| 6 | 2 | 4 | 6 | 0 | 6 | 2338.0121 | 2338.0128 | -0.0008 |
| 7 | 2 | 5 | 7 | 0 | 7 | 2338.0194 | 2338.0194 | 0.0000 |
| 4 | 3 | 1 | 4 | 1 | 3 | 2338.0229 | 2338.0225 | 0.0004 |
| 6 | 1 | 5 | 6 | 2 | 4 | 2338.1135 | 2338.1132 | 0.0003 |
| 3 | 3 | 1 | 3 | 2 | 2 | 2338.2318 | 2338.2317 | 0.0001 |
| 6 | 2 | 5 | 6 | 1 | 6 | 2338.2677 | 2338.2679 | -0.0002 |
| 2 | 2 | 0 | 2 | 0 | 2 | 2338.3443 | 2338.3442 | 0.0001 |
| 7 | 2 | 6 | 7 | 1 | 7 | 2338.3621 | 2338.3610 | 0.0011 |

```
 2  2  1    1  0  1 2338.5812  2338.5819  -0.0006

 8  2  7    7  2  5 2339.1111  2339.1103   0.0008

 6  3  4    5  1  4 2339.1309  2339.1313  -0.0005
```

******************************************************************